\documentclass[aip,jcp,preprint,floatfix,amsmath,amsfonts,14pt]{revtex4-1}
\usepackage[top=2cm,bottom=2cm,left=2.5cm,right=2.5cm]{geometry}

\usepackage{setspace}

\usepackage{graphicx}
\usepackage{amsmath}
\usepackage[utf8]{inputenc}
\usepackage{color}
\usepackage{bbold}

\usepackage{upgreek}

\usepackage[normalem]{ulem}

\usepackage{subfigure}
\usepackage{braket}

\begin{document}

\title{{Towards a fully size-consistent method of increments}}
\author{E. Fertitta}
\author{D. Koch}
\author{B. Paulus}
\affiliation{Institut f\"ur Chemie und Biochemie - Freie Universit\"at Berlin,\\
Takustr. 3, 14195 Berlin, Germany}
\author{G. Barcza}
\author{\"O. Legeza}
\affiliation{Strongly correlated systems ``Lend\"ulet'' research group,\\ Wigner Research Centre for Physics, P.O. Box 49, H-1525 Budapest, Hungary}

\date\today

\begin{abstract}
The method of increments (MoI) allows one to successfully calculate cohesive energies of bulk materials with high accuracy, but it encounters difficulties when calculating whole dissociation curves. The reason is that its standard formalism is based on a single Hartree-Fock (HF) configuration whose orbitals are localized and used for the many-body expansion. Therefore, in those situations where HF does not allow a size-consistent description of the dissociation, the MoI cannot yield proper results either. Herein we address the problem by employing a size-consistent multiconfigurational reference for the MoI formalism. This leads to a matrix equation where a coupling derived by the reference itself is employed. In principle, such approach allows one to evaluate approximate values for the ground as well as excited states energies. While the latter are accurate close to the avoided crossing only, the ground state results are very promising for the whole dissociation curve, as shown by the comparison with density matrix renormalization group (DMRG) benchmarks. We tested this two-state constant-coupling (TSCC)-MoI on beryllium rings of different sizes and studied the error introduced by the constant coupling.\\
\end{abstract}

\maketitle

\vspace{1cm}
\section{Introduction}

In the past few decades, the development of local correlation methods has given the chance to perform accurate calculations mainly on ground state properties for a variety of extended and periodic systems. The success of such methods is given by the fact that they constitute, in many cases, the only valid alternative to density functional theory (DFT)\cite{Dreizler1990, Eschrig1996, Kohn1965}, apart maybe from more sophisticated approaches such as the ab initio density matrix renormalization group (DMRG)\cite{White1992,White1993,Legeza2008,Chan2009,Marti2010,Schollwock2011,Wouters-2014e,Olivares-Amaya-2015,Murg2015,Szalay-2015} and stochastic methods.\cite{Scuseria2008,Booth2009,Spencer2012,Morales2012,Cleland2012,Shepherd2012,Kolodrubetz2012,Petruzielo2012,Roggero2013,Willow2014}\\
In the framework of local correlation approaches\cite{Pulay1983,Pulay1986,Stollhoff1991,Pardon1995,Kitaura1999,Schutz2000,Schutz2001,Schutz2002,Schutz2002a,Schutz2014,Schwilk2015,Pisani2008,Pisani2012}, the method of increments (MoI) was first developed by Stoll\cite{Stoll1992a,Stoll1992b,Stoll2009,Stoll2010} in the 1990s and further applied by other groups on extended and periodic systems\cite{Paulus2003,Paulus2006,Voloshina2007,Schmitt2009,Muller2011,Friedrich2007,Friedrich2008,Friedrich2012}. The MoI is a powerful and flexible method which allows the application of different formalisms and yields accurate results for ground state properties of crystals including strongly correlated bulk metals\cite{Paulus2003a,Alsheimer2004,Voloshina2012,Voloshina2014}. For the latter, the use of multiconfigurational (MC) and multireference (MR) formalisms of the MoI are necessary, but they require a proper choice of the localization scheme. In recent works\cite{Fertitta2015,Koch2016}, we have tested and analyzed the flexibility of the MoI, applying a complete active space (CAS) and a MR formalism for the calculation of the correlation energy in a strongly correlated model system, \textit{i.e.} beryllium rings. Despite the fact that the methods allowed us to retrieve significant amount of the correlation energy over the whole dissociation curve, we were not able to describe the behavior at the avoided crossing with sufficient accuracy . This was the case because the localized orbitals (LOs) used in the MoI were generated by unitary transformation from a single Hartree-Fock (HF) configuration which was dominant in a particular region of the dissociation curve. In the crossing region, where the multiconfigurational character increases sharply, this starting set of LOs was obviously not sufficient. In the present work, we present a solution to this problem by coupling the results achievable from both configurations. Formally, this allows one to achieve a fully size-consistent formalism of the method of increments which can be then applied for calculating fragmentation processes, rather than dissociation/binding energies, by the calculation of a continuous ground state dissociation curve in all distance regimes. It has to be underlined that, the concept of size-consistency is often used interchangeably with the term of size-extensivity in literature. However, while the latter refers to the correct scaling with the particle number, size-consistency refers to the requirement of correctly describing fragmentation when two fragments are moved apart. This is not a property of the method itself, but depends rather on the process under study. If multireference methods are employed, this requirement can be satisfied by employing a proper size-consistent reference. In the same way, the approach proposed herein aims to solve the size-consistency lack of the MoI by using a MC reference rather than a HF reference.\\ 
This paper is structured as follows: in section~\ref{sec_comp_det} the computational details are described; the MoI general formalism is briefly summarized in section~\ref{sec_moi} together with the new two-state formalism; in section~\ref{sec_results} we report the results obtained for our model system, Be$_6$ ring, and compare them to DMRG benchmarks; also the results obtained for larger beryllium systems are discussed; we finally draw our conclusion in section~\ref{sec_conclusion}. 

\section{Computational details}\label{sec_comp_det}

All calculations presented in this work, including the Foster-Boys localizations\cite{Foster1960} and MoI calculations, were performed employing the quantum chemical program package MOLPRO\cite{MOLPRO}. The dissociation curves of the different ring-shaped beryllium clusters are reported as a function of the Be-Be internuclear distance which was varied imposing the condition of being equal all over the system. This way the $D_{nh}$ symmetry is always ensured. The results presented in this work were obtained using the $cc$-pVDZ basis set\cite{Prascher2011} and a minimal basis set $(9s,4p) \rightarrow [2s,1p]$ derived by its contraction. It has to be underlined that, although MOLPRO can handle Abelian groups only (in our case $D_{2h}$), we will refer to irreducible representations of the full point group $D_{nh}$ to label states and orbitals.\\
Owing to the multiconfigurational nature of the problem where even the smallest test cases are beyond the feasibility of standard CASSCF approaches, we applied the ab initio version of the DMRG method to benchmark MoI results. In the DMRG calculations, besides the ground state the first excited $^1A_{1g}$ state of the Be$_6$ ring was targeted. For each Be-Be distance, we performed DMRG calculations using the QCDMRG-Budapest program package\cite{qcdmrg-budapest}. The number of block states were chosen by the dynamical block state selection (DBSS) approach\cite{Legeza2003,Legeza2004} with an apriory set value of the quantum information loss $\chi=10^{-4}$ (maximum block states up to 6000). The initialization of the DMRG was optimized using the configuration interaction based dynamically extended active space (CI-DEAS) procedure\cite{Legeza2003a}.

\section{The method of increments}\label{sec_moi}

\subsection{Standard formalism}\label{sec_moi_general}

Exploiting the short range nature of the electron correlation, the method of increments employs localized orbitals (LOs) as an orbital basis for correlation calculations. These are obtained by an unitary transformation of the canonical orbitals of a reference HF wave function and then collected according to their location in space into different groups, referred to as bodies. By doing so, this approach aims to describe the correlation energy $E_{\rm corr}$  as a sum of individual contributions (increments) associated to different parts of the system.\\
Starting from the crudest approximation, that is the sum of one-body increments $\epsilon_{\rm i}$, one can improve the description of $E_{\rm corr}$ step by step by including higher-order increments. These are calculated considering the correlation of all pairs (2-body increments), triples (3-body increments), quadruples (4-body increments) of bodies and so on. As an instance, the 2-body increments $\Delta \epsilon_{\rm ij}$ can be calculated as:
\begin{equation}
\Delta \epsilon_{\rm ij} = \epsilon_{\rm ij} - \left(\epsilon_{\rm i} + \epsilon_{\rm j}\right)\label{eq_2-body}
\end{equation}
where $\epsilon_{\rm ij}$ is the correlation energy for the pair of bodies $i,j$. Similarly, one can calculate 3-body increments $\Delta \epsilon_{\rm ijk}$:
\begin{equation}
\Delta \epsilon_{\rm ijk} = \epsilon_{\rm ijk} - \left(\Delta \epsilon_{\rm ij} + \Delta \epsilon_{\rm jk} + \Delta \epsilon_{\rm ik}\right) - \left(\epsilon_{\rm i} + \epsilon_{\rm j} + \epsilon_{\rm k}\right)\label{3-body}
\end{equation}
or higher order contributions in an analogous way.\\
Finally, these contributions can be summed up yielding the total correlation energy:
\begin{equation}
    E_{\rm corr} = \sum_{\rm i} \epsilon_{\rm i} + \sum_{\rm i<j} \Delta \epsilon_{\rm ij} + \sum_{\rm i<j<k} \Delta \epsilon_{\rm ijk} + \cdots\label{eq_exp_corr}
\end{equation}
The choice of the bodies is arbitrary, but an insight into the electronic structure of the system might help to make a more proper partitioning which allows faster convergence. Independently of these choices, the MoI will bring a substantial advantage only if the expansion in Eq.~\ref{eq_exp_corr} can be truncated. In practice, this can be done if the increments converge with distance between the involved bodies and with the order.\\
Depending on the quantum chemical method of choice, different sets of orbitals have to be unitarily transformed in order to get appropriate localized orbital subspaces. For instance in order to employ a single-reference method such as coupled cluster (CC), only the valence orbitals of the HF reference are localized and grouped into bodies, and the excitations are performed into the delocalized virtual orbitals. On the other hand, in order to apply multireferece (MR) methods a different scheme is used which requires multiconfigurational MoI calculations performed using both localized valence and virtual orbitals which are grouped together into bodies. This CAS-MoI approach allows one to calculate static correlation contributions while dynamical correlation can then be calculated including the remaining delocalized virtual orbitals in a MR calculation.\\
The MR-MoI was shown to be particularly successful to describe situations where local static correlation plays an important role, for example for calculating the cohesive energy of bulk alkaline earth metals.\cite{Voloshina2014} Moreover, we tested recently the behavior of this method for the description of the whole dissociation curve a one-dimensional beryllium system, where an avoided crossing was observed. Also in this case, the MR-MoI yielded very good results, but we were limited by the fact that in different regions of the dissociation curve different HF configurations had to be selected for the localization in order to obtain reasonably converging results. Indeed, it is important to underline that the inclusion of static correlation via this procedure cannot compensate for the absence of size-consistency because the MC wave functions involve local excitations by employing an LO basis constructed by a single reference.

\subsection{The two-state constant-coupling MoI}\label{sec_tscc_moi}

\begin{figure}[h]
\includegraphics[width=0.4\textwidth]{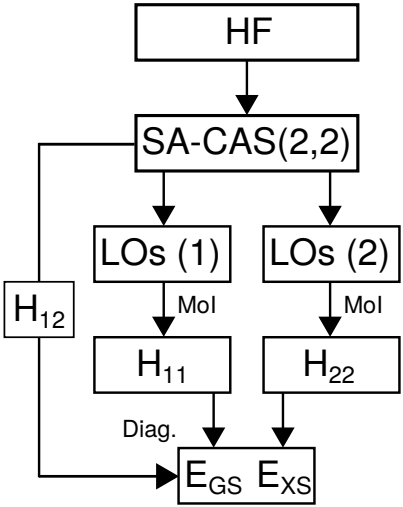}
\caption{(Color online) Schematic representation of the two-state constant-coupling MoI as employed for describing the dissociation of beryllium rings. After performing a state-average CAS(2,2) calculation on top of the Hartree-Fock wave function, the natural orbitals of the two states are used for localization. This allows one to obtain two sets of localized orbitals which are used separately for the method of increments calculations yielding the diagonal elements of the $2\times2$ Hamiltonian matrix. The off-diagonal elements are approximated using the coupling obtained from the state-average CAS(2,2). By diagonalization the eigenvalues for the ground and excited states can be calculated.}
\label{fig_1}
\end{figure}

In this section we present our solution to the lack of size-consistency of the MoI. Since, as explained so far, the problem lies in the choice of the reference consisting of a single configuration, we decided to start from a MC reference instead. Consider, for instance, a system where two configurations $\ket{{\phi}^{'}}$ and $\ket{{\phi}^{''}}$, with respective energies $E^{'}$ and $E^{''}$, are dominant in different regions of the dissociation curve. Trivially, a proper size-consistent reference for the ground state wave function will be given by:
\begin{equation}
    \ket{\tilde{\Phi}_{\rm GS}} = c^{'}_{\rm GS}\ket{{\phi}^{'}} + c^{''}_{\rm GS}\ket{{\phi}^{''}}\label{ref_two_state}
\end{equation}
After localizing the orbitals constituting these two configurations, one can use them separately as basis for MoI calculations in an analogous way to what was described in section~\ref{sec_moi_general}. This way one can evaluate two different sets of increments ($\epsilon^{'}_{\rm i}$, $\Delta\epsilon^{'}_{\rm ij}$, $\Delta\epsilon^{'}_{\rm ijk}$, etc. from $\ket{\phi^{'}}$ and $\epsilon^{''}_{\rm i}$, $\Delta\epsilon^{''}_{\rm ij}$, $\Delta\epsilon^{''}_{\rm ijk}$, etc. from $\ket{\phi^{''}}$) and use them as correlation corrections for $E^{'}$ and $E^{''}$ using a similar expression to Eq.~\ref{eq_exp_corr}:
\begin{eqnarray}
    E^{'}_{\rm corr} & =& \sum_{\rm i} \epsilon^{'}_{\rm i} + \sum_{\rm i<j} \Delta\epsilon^{'}_{\rm ij} + \sum_{\rm i<j<k} \Delta\epsilon^{'}_{\rm ijk} + \cdots \label{eq_moi_a}\\
    E^{''}_{\rm corr} & =& \sum_{\rm i} \epsilon^{''}_{\rm i} + \sum_{\rm i<j} \Delta\epsilon^{''}_{\rm ij} + \sum_{\rm i<j<k} \Delta\epsilon^{''}_{\rm ijk} + \cdots \label{eq_moi_b}
\end{eqnarray}
The electron-correlation corrected terms $E^{'} + E^{'}_{\rm corr}$ and $E^{''} + E^{''}_{\rm corr}$ constitute the diagonal elements of a $2\times 2$ Hamiltonian matrix with the corresponding secular equation
\begin{equation}
\begin{vmatrix}
H_{11}-E  & H_{12} \\
H_{21}  & H_{22}-E 
\end{vmatrix}=0
\hspace{0.7cm} \mbox{with} \hspace{0.7cm}
\begin{cases}
H_{11}=E^{'} + E^{'}_{\rm corr}\\
H_{22}=E^{''} + E^{''}_{\rm corr}
\end{cases}
\label{H_mat}
\end{equation}
Eq.~\ref{H_mat} implies two orthonormal bases $\ket{\upphi_1}$ and $\ket{\upphi_2}$ which include correlation since they are constructed by local excitations from $\ket{{\phi}^{'}}$ and $\ket{{\phi}^{''}}$, respectively. However, since the MoI yields corrections to the energy, but does not deal directly with the wave function, the calculation of the coupling term $H_{12}=\bra{{\upphi_{1}}}\hat{\mathcal H}\ket{{\upphi_{2}}}$ is something we cannot straightforwardly achieve via this approach. In order to overcome this problem, we chose to neglect the corrections necessary for describing the full correlated system and to use the value $\bra{{\phi^{'}}}\hat{\mathcal H}\ket{{\phi^{''}}}$  for $H_{12}$ instead. We will see in the discussion of our results that this approximation yields reasonable results for the system  under study.\\
By solving the secular equation (Eq.~\ref{H_mat}), two energy values will be obtained, $E_{\rm GS}$ and $E_{\rm XS}$, corresponding to the ground and first excited state wave function, respectively. Even if our main interest is to calculate the ground state energy in a size-consistent manner, the chance of obtaining information about excited states via the MoI is, of course, appealing. We will therefore discuss the corresponding results as well. However, as we will see, while the ground state energies are in very good agreement with our DMRG benchmarks, the excited state energies show larger deviations.\\ 
It should be pointed out that, despite not strictly required, in order to allow a more straightforward MoI treatment the reference configurations $\ket{\phi^{'}}$ and $\ket{\phi^{''}}$ should be closed shell and their orbitals should be easily localizable. In the system under investigation these conditions are fulfilled. A meaningful approach to obtain the required references is to perform a state-averaged (SA) CAS-SCF calculation with a proper active space. By applying a SA approach, we can calculate a first approximation to both the ground and excited state in terms of the two main configurations. At this point, as already stated, two different unitary transformations, one for $\ket{\phi^{'}}$ and the other for $\ket{\phi^{''}}$, can be applied yielding two sets of LOs that can be used for the MoI calculations, separately. We will refer to the approach described in this section as two-state constant-coupling MoI (TSCC-MoI) which is schematized in Fig.~\ref{fig_1} as employed in this work.\\
Since each of the two diagonal elements $H_{11}$ and $H_{22}$ relies on a single reference ($\ket{\phi^{'}}$ and $\ket{\phi^{''}}$), they can be calculated by applying any MoI formalism discussed so far, including CCSD(T)-MoI, CAS-MoI and MRCI-MoI. According to the method employed in this step we distinguish CCSD(T)-, CAS-, MRCISD-, MRCISD(+Q)-, CAS-CI-TSCC-MoI. The latter is analogous to the CAS formalism described so far, with the difference that no orbital optimization is performed. 

\section{Results}\label{sec_results}

\subsection{Be$_{6}$ Ring}

\begin{figure}[h]
\includegraphics[width=0.92\textwidth]{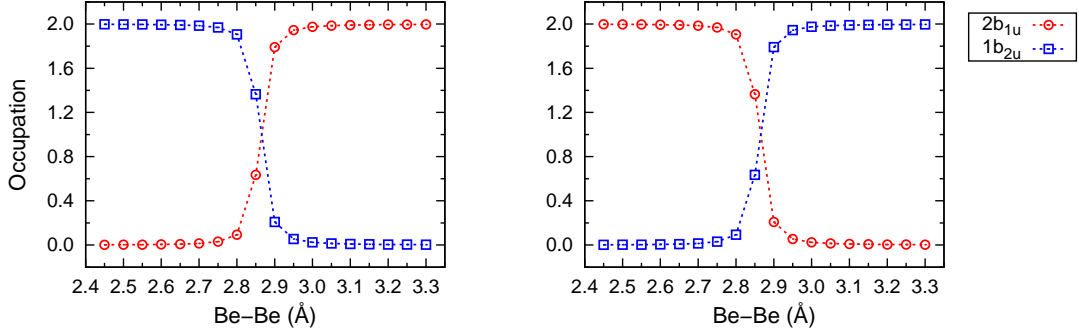}
\caption{(Color online) Occupation number of the $2b_{1u}$ and $1b_{2u}$ orbitals of the Be$_6$ ring as calculated with a minimal $(9s,4p)\rightarrow[2s,1p]$ basis set via SA-CAS(2,2). The left and right panel show the values for the ground and first excited $^1A_{1g}$ state, respectively. In both states, the orbitals $2a_{1g}$, $2e_{1u}$ and $2e_{2g}$ are fully occupied.}
\label{fig_2}
\end{figure}
\begin{figure*}[h]
\centering
\includegraphics[width=0.8\textwidth]{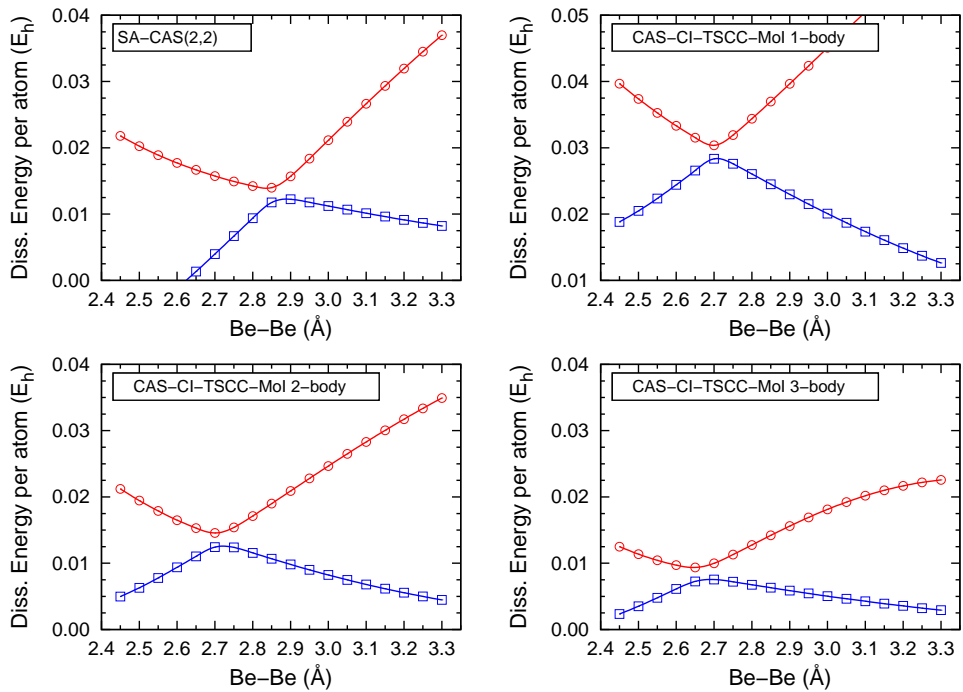}
\caption{(Color online) Ground (blue lines) and first excited (red lines) $^1A_{1g}$ state dissociation curves of Be$_6$ ring as obtained via the two-state constant-coupling MoI at the CAS-CI level employing a minimal $(9s,4p) \rightarrow [2s,1p]$ basis set. In the upper left panel, the reference energies obtained via SA-CAS(2,2) calculations are shown. The results obtained by applying the one-body, two-body and three-body corrections are also shown. In all cases energies per atom are reported.}\label{fig_3}
\end{figure*}
\begin{figure}[htb]
\centerline{
\includegraphics[width=0.4\textwidth]{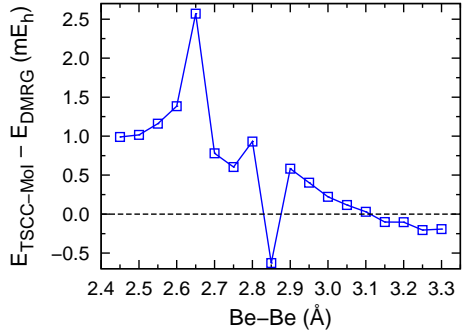}
}
\caption{(Color online) Difference between the ground state energies of the Be$_6$ ring calculated via CAS-CI-TSCC-MoI up to the 3-body level ($E_{\rm TSCC-MoI}$) and DMRG with $\chi = 10^{-4}$ ($E_{\rm DMRG}$) with a minimal $(9s,4p) \rightarrow [2s,1p]$ basis set.}\label{fig_4}
\end{figure}

By applying the standard MoI to the ground state dissociation curve of the Be$_{6}$ ring, a proper description of the avoided crossing with the first excited $^1A_{1g}$ state cannot be achieved. As discussed, this is caused by the lack of size-consistency of the HF reference since two configurations related by double excitations play a dominant role in different regions of the dissociation curve. This can be seen in Fig.~\ref{fig_2} where we report the occupation numbers of the frontier orbitals ($2b_{1u}$ and $1b_{2u}$) of the Be$_6$ ring as calculated with a SA-CAS(2,2) and a minimal basis set. Data for both the ground and first excited $^1A_{1g}$ state are reported. For internuclear distances shorter than 2.80~{\AA}, the ground state wave function is clearly dominated by the configuration $2a_{1g}^22e_{1u}^42e_{2g}^41b_{2u}^2$ ($Conf1$) while above 2.90~{\AA}, the configuration $2a_{1g}^22e_{1u}^42e_{2g}^42b_{1u}^2$ ($Conf2$) is predominant. For the excited state, the situation is opposite to the ground state. In the two extreme situations where one of the two configurations is dominant, the standard MoI can be applied by localizing the orbitals of $Conf1$ or $Conf2$. However, since around the crossing (2.85~\AA) the contribution of both configuration is comparable, the standard approach cannot be easily applied. Then it is necessary to couple the results yield by MoI for $Conf1$ and $Conf2$ via the TSCC-MoI. We started by testing this method with a minimal basis set and using the SA-CAS(2,2) described above as reference. The localization of the virtual orbitals was performed separately for the orbitals symmetric and antisymmetric with respect to the plane of the ring $\sigma_h$, yielding six $p_z$-like LOs (where the $z$-axis is perpendicular to the $\sigma_h$) and twelve localized orbitals with mixed $sp$ character. Taking into account the six occupied valence orbitals, this gives rise to four LOs for each body, each including two active electrons. This way at the CAS-CI-TSCC-MoI or CAS-TSCC-MoI, the one-, two- and three-body corrections are obtained via CAS(2,4), CAS(4,8) and CAS(6,12) calculations, respectively. This clearly constitutes a drastic reduction of the active space with respect to the full valence CAS(12,24). The CAS-CI-TSCC-MoI results obtained with the $(9s,4p)\rightarrow[2s,1p]$ basis set are reported in Fig.~\ref{fig_3}. As it can be seen, starting from the SA-CAS(2,2) reference and applying the one-body, two-body and three-body corrections, the avoided crossing is always found and its position shifts towards smaller interatomic distances as correlation is included. The energy difference between the CAS-CI-TSCC-MoI results and the DMRG with DBSS benchmark for the ground state are reported  in Fig.~\ref{fig_4}. As one can see, the agreement between the two approaches is very good for the ground state, since the energy difference between the two methods lies in the range -0.5--2.0~${\rm m}E_h$ around the avoided crossing.\\

\begin{figure}[h]
    \includegraphics[width=0.92\textwidth]{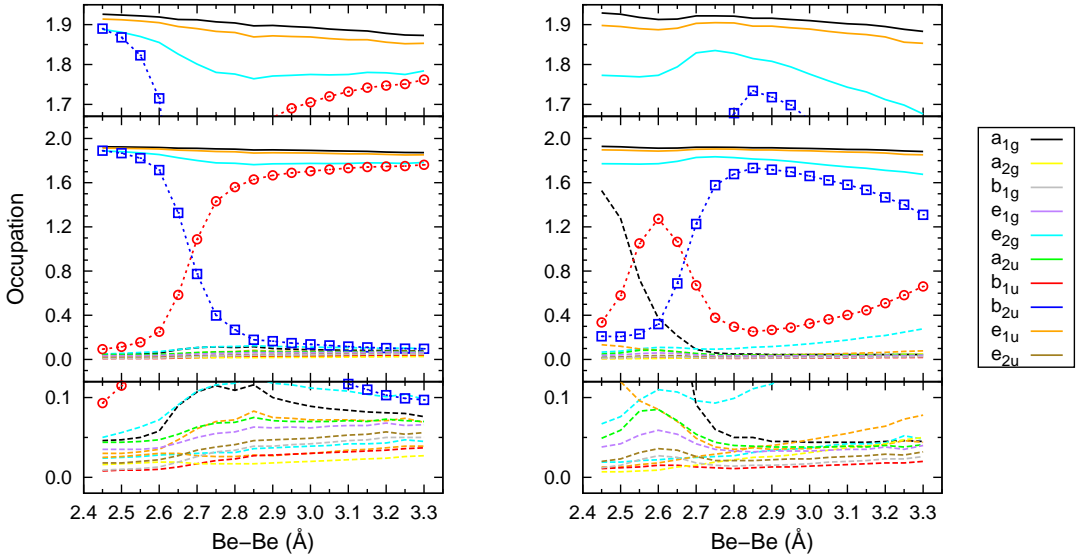}
    \caption{(Color online) Occupation number of the 24 valence orbitals of the Be$_6$ ring as calculated with a minimal $(9s,4p)\rightarrow[2s,1p]$ basis set via DMRG. The left and right panel show the values for the ground and first excited $^1A_{1g}$ state, respectively. Note that the legend refers to the symmetry of the orbitals in $D_{6h}$, rather than to the individual orbitals.	Solid and dashed lines indicate orbitals which in the SA-CAS(2,2) reference are fully occupied and empty, respectively. Dotted lines refer to $2b_{1u}$ (circles) and $1b_{2u}$ (squares).}
\label{fig_5}
\end{figure}

For the excited state, the TSCC-MoI yields a acceptable accuracy ($1-5$~m$E_h$) only around the avoided crossing (between 2.60{\AA} and 2.80{\AA}), but in other distance regimes the difference with respect to DMRG results rises over $10~{\rm m}E_h$. This can be explained by the fact that the employed reference does not provide a description for the excited state as good as for the ground state. The SA-CAS(2,2) was indeed designed to include the two configurations important for the ground state, but a different reference might be more suitable for this excited state. This can be seen in Fig.~\ref{fig_5} where we report the occupation numbers of the 24 valence orbitals (accounting for the $2s$ and $2p$ orbitals for each Be atom) of the Be$_6$ ring as obtained by DMRG calculations. When comparing these data with the ones presented in Fig.~\ref{fig_2}, the difference between the ground and excited state is stricking. Despite a shift of the crossing, the ground state presents a similar wave function to the SA-CAS(2,2) reference, since also the DMRG wave function is clearly dominated by $Conf1$ and/or $Conf2$ depending on the internuclear distance regime. This is not the case for the excited state where the multireference character is much more pronounced according to DMRG results. In fact, the occupation patterns revail that the wave function is dominated by the reference configurations, $Conf1$ and $Conf2$, only between 2.60~{\AA} and 2.80~{\AA}, while in other distance regimes the deviation between the occupation of the reference and the DMRG wave function are significant. For instance, for Be-Be distances shorter than 2.60~{\AA} the orbital $3a_{1g}$ competes with $2b_{1u}$. For large interatomic distances this picture can be understood considering that the dissociation limit for the first excited $^1A_{1g}$ state is constituted by four beryllium atoms in $^1S_g$ state and two in a $^3P_u$ state. Their combinations in an extended system yield a wave function with very strong multireference character.\\
Without forgetting the limited accuracy of the excited state far from the crossing region, we will show in the following TSCC-MoI results for both states. Moreover, we will use them to discuss excitation energies at the avoided crossing, where we have shown and justified that both state have an acceptable accuracy.\\

\begin{figure}
\includegraphics[width=0.4\textwidth]{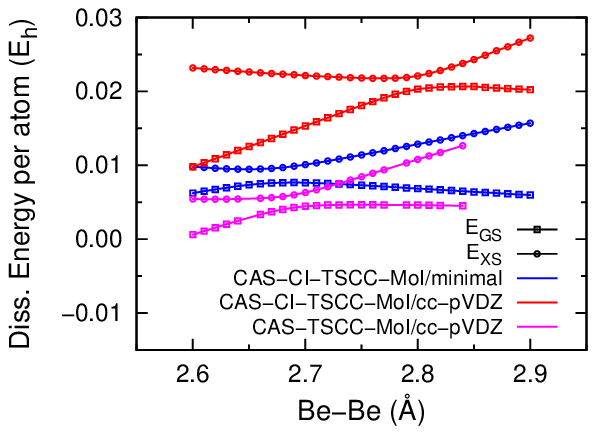}
\includegraphics[width=0.4\textwidth]{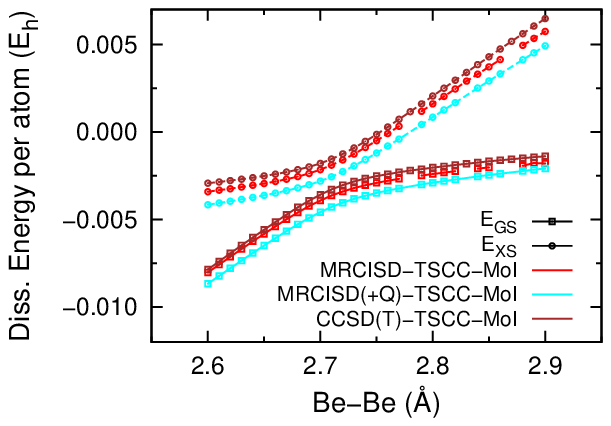}
\caption{(Color online) Dissociation curves of the ground and first excited $^1A_{1g}$ state of Be$_6$ ring around the crossing region as obtained with the two-state constant-coupling method of increments (TSCC-MoI) and different basis sets. On the left panel, CAS-CI-TSCC-MoI with minimal and $cc$-pVDZ basis set and CAS-TSCC-MoI with $cc$-pVDZ. On the right, MRCISD-, MRCISD(+Q)- and CCSD(T)-TSCC-MoI with $cc$-pVDZ.  In all cases energies per atom are reported.}
\label{fig_6}
\end{figure}

In order to improve the description of the system we used a better basis set, namely $cc$-pVDZ, and included dynamical correlation. The first problem to face when using a larger basis set is the localization of the virtual orbitals if CAS-TSCC-MoI or CAS-CI-TSCC-MoI is applied. Indeed, while for the minimal basis set the selection of the orbitals that have to be localized is rather obvious, this is not in general the case for a larger basis set. In the case of $cc$-pVDZ we proceeded as follows in order to obtain meaningful virtual LOs to be used for the calculations. As done for the minimal basis set, firstly we exploit symmetry in order to select two groups of orbitals to be localized. This way the lowest-energy six molecular orbitals antisymmetric with respect to $\sigma_h$ were localized yielding LOs with mostly $p_z$ contribution. Then all the virtual orbitals lying on the plane of the ring were localized and among the resulting LOs only the lowest-energy twelve were selected. These virtual LOs together with the valence localized orbitals were then used for CAS-TSCC-MoI calculations as described above. The orbital optimization occuring during CAS-(TSCC)-MoI calculations redelocalizes the orbitals not included for the construction of the active space. These will then be included in the MRCI step for the calculation of dynamical correlation.\\
In the left panel of Fig.~\ref{fig_6} the CAS-CI- and CAS-TSCC-MoI results obtained with $cc$-pVDZ are compared with the CAS-CI-TSCC-MoI data for the minimal basis set. The first striking fact is that the dissociation energy calculated with the minimal basis set is closer to the CAS-TSCC-MoI/$cc$-pVDZ than the CAS-CI-TSCC-MoI/$cc$-pVDZ. This observation can be easily explained by considering that with a minimal basis set, the CAS-CI-TSCC-MoI converges towards the Full-CI limit as the incremental order increases and the orbital optimization does not play such an important role if sufficient increments are included. As one can see, this is instead crucial for the VDZ basis set and has a major effect on the position of the avoided crossing, which is shifted to larger internuclear distances if orbital optimization is not considered. This was quite expected and highlights once again the importance of static correlation for this system.\\
In order to include the dynamical correlation, the method was also applied at the MRCISD-, MRCISD(+Q)- and CCSD(T)-TSCC-MoI level. The right panel of Fig.~\ref{fig_6} shows the resulting dissociation curves at the three-body level. As one can see, the energy difference between CCSD(T)-TSCC-MoI and the MR approaches is very small (in the order of $5~{\rm m}E_h$) and also the position of the avoided crossing is not very sensible to this choice. We find this particularly important, since obviously the application of the CCSD(T) scheme is easier than the MR one since it does not require to localize the virtual orbitals, as previously explained.\\
In Table~\ref{tab_1} we report the energy gap at the avoided crossing, $\Delta E_\mathrm{Cr}$, and the corresponding interatomic distance $R_\mathrm{Cr}$ as calculated with the TSCC-MoI at different levels of theory for $cc$-pVDZ as well as for CAS-CI-TSCC-MoI with the minimal basis set. As one can see, orbital optimization causes a large shift to smaller bond distances from CAS-CI- to CAS-TSCC-MoI with $cc$-pVDZ which is corrected to slightly larger bond lengths by the incorporation of dynamical correlation. The latter also leads to a slightly increased gap size, which is lower for MRCISD(+Q)- and CCSD(T)-TSCC-MoI than for MRCI-TSCC-MoI energies. Furthermore, there is a relatively large decrease in $\Delta E_\mathrm{Cr}$ with $cc$-pVDZ with respect to the minimal basis set already at the CAS level, probably caused by the higher flexibility of the valence-double-zeta basis set. On the other hand, the avoided crossing position obtained with the minimal basis set is in rather good agreement with the one from $cc$-pVDZ. Clearly the energy gap is strongly related to the constant coupling employed being $\Delta E_\mathrm{Cr} \approx 2|\bra{{\phi^{'}}}\hat{\mathcal H}\ket{{\phi^{''}}}|$ as shown in Table~\ref{tab_1}. Therefore the values for the gap evaluated in such a way have to be taken with care because $\bra{{\phi^{'}}}\hat{\mathcal H}\ket{{\phi^{''}}}$ includes no correlation.\\
We conclude this section by comparing the results obtained with the conventional MoI and the TSCC-MoI. As already stated, in our previous works\cite{Fertitta2015,Koch2016} we discussed how two different configurations have to be employed for different internuclear distance regimes and how the dissociation curves obtained from these configurations cross. This is highlighted in Fig.~\ref{fig_7} where we compare the dissociation curves obtained via CCSD(T)-MoI and CCSD(T)-TSCC-MoI. In the supplementary materials similar comparisons for other methods are also shown. The energy difference of the two curves for each state are also shown. $\Delta E_\mathrm{GS}$ denotes the energy difference between the ground state of the TSCC-MoI and the lowest energy obtained from the two MoI calculations. On the other hand, the difference between the TSCC-MoI excited state and the higher MoI energy is labeled with $\Delta E_\mathrm{XS}$. As expected, the differences are largest around the crossing point. Moreover, the differences are smaller for short atomic distances than for the larger ones. As one can see, the energy difference of the excited state to the higher MoI energy is larger than the difference between ground state and lower MoI energy.
\begin{figure*}
\subfigure{\includegraphics[width=0.4\textwidth]{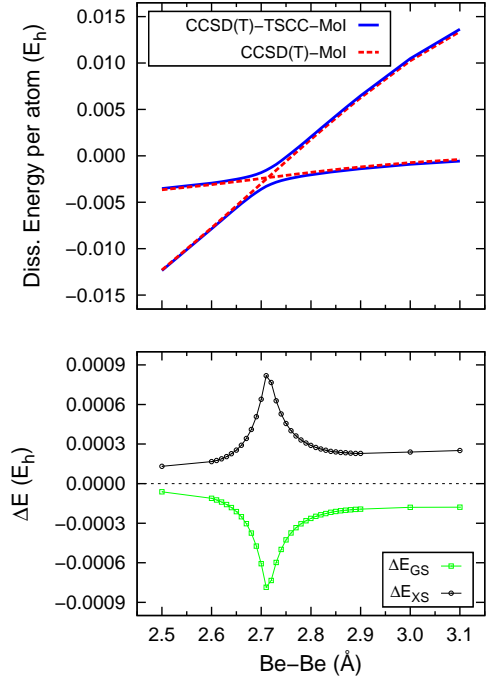}}
\caption{(Color online) Comparison between the two-state constant-coupling method of increments (TSCC-MoI) results and the ones obtained with the standard method of increments for the Be$_6$ ring. In both  cases CCSD(T) was employed at the three-body level with the $cc$-pVDZ basis set. The reported values are energies per atom.}
\label{fig_7}
\end{figure*} 

\begin{table}[h]
\centering
\caption{Smallest energy differences between the ground and first excited $^1A_{1g}$ state, $\Delta E_\mathrm{Cr}$ (in m$E_h$), and the corresponding interatomic distances, $R_\mathrm{Cr}$ (in \AA), as obtained with the two-state constant-coupling MoI with different methods and the $cc$-pVDZ basis set at a three-body level. For comparison also the values $2|\bra{{\phi^{'}}}\hat{\mathcal H}\ket{{\phi^{''}}}|$ (in m$E_h$) are reported.}
\begin{tabular}{llp{0.6cm}cp{0.6cm}cp{0.6cm}c}
\\
Basis set & Method && $R_\mathrm{Cr}$ && $\Delta E_\mathrm{Cr}$  && $2|\bra{{\phi^{'}}}\hat{\mathcal H}\ket{{\phi^{''}}}|$  \\ \hline 
$(9s,4p)\rightarrow[2s,1p]$ & SA-CAS(2,2)   & & 2.88 & & 14.852  & & 12.571 \\
          & CAS-CI-TSCC-MoI                 & & 2.67 & & 11.812  & & 11.792 \\
$cc$-pVDZ & SA-CAS(2,2)                     & & 2.84 & & 11.078  & & 10.907  \\
          & CAS-CI-TSCC-MoI                 & & 2.80 & & 10.766  & & 10.761 \\
          & CAS-TSCC-MoI                    & & 2.69 & & 10.403  & & 10.327 \\
          & MRCISD-TSCC-MoI                 & & 2.71 & & 10.416  & & 10.408 \\ 
          & MRCISD(+Q)-TSCC-MoI             & & 2.71 & & 10.414  & & 10.408 \\
          & CCSD(T)-TSCC-MoI                & & 2.71 & & 10.413  & & 10.408
\end{tabular}
\label{tab_1}
\end{table}

\subsection{On the validity of the constant coupling}\label{sec_coupling}

\begin{figure*}[h]
    \centerline{
    \includegraphics[width=0.8\textwidth]{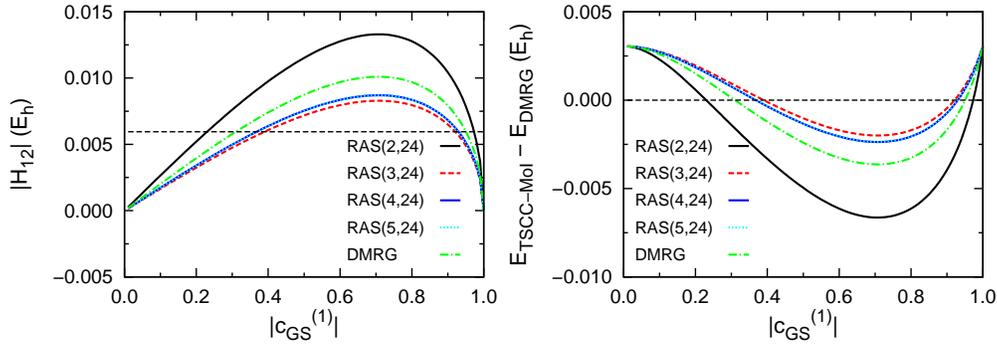}
}
\caption{(Color online) On the left, dependence of the coupling $H_{12}$ on the coefficient $c^{(1)}_{\rm GS}$ as described by Eq.~\ref{eq_coupling_gap}. The gap $E_{\rm XS}-E_{\rm GS}$ was calculated employing different RAS-SCF methods and DMRG with $\chi=10^{-4}$. The dashed line indicates the coupling calculated by the SA-CAS(2,2) reference. On the right, differences between CAS-CI-TSCC-MoI ground state energies ($E_{\rm TSCC-MoI}$) obtained with these values for $H_{12}$ and the DMRG ground state energy ($E_{\rm DMRG}$). A minimal $(9s,4p)\rightarrow[2s,1p]$ was employed and all values refer to an internuclear distance of 2.70~\AA.}
\label{fig_8}
\end{figure*}

As discussed, within the TSCC the diagonal elements of the Hamiltonian matrix are evaluated using the two SA-CAS(2,2) configurations as reference for a many-body expansion. However, the main assumption of the introduced formalism is that the off-diagonal elements can be kept constant for different levels of accuracy. This is a strong approximation whose validity depends on the system under study and needs to be tested. In this section, we will illustrate some arguments concerning the evaluation of a better coupling term than the one obtained by the employed reference and discuss how this affects the final result.\\
Consider the two basis functions implied by Eq.~\ref{H_mat}, $\ket{\upphi_1}$ and $\ket{\upphi_2}$. They can be expressed as a linear combination of the ground and excited state wave functions $\ket{\Phi_{\rm GS}}$ and $\ket{\Phi_{\rm XS}}$:
\begin{eqnarray}
    \ket{\upphi_1} &=& c^{(1)}_{\rm GS} \ket{\Phi_{\rm GS}} + c^{(1)}_{\rm XS} \ket{\Phi_{\rm XS}} \\
    \ket{\upphi_2} &=& c^{(2)}_{\rm GS} \ket{\Phi_{\rm GS}} + c^{(2)}_{\rm XS} \ket{\Phi_{\rm XS}}
\end{eqnarray}
It follows then that the coupling term $H_{12} = \bra{{\upphi_{1}}}\hat{\mathcal H}\ket{{\upphi_{2}}}$ can be written as:
\begin{eqnarray}
    |H_{12}| &=& c^{(1)}_{\rm GS}c^{(2)}_{\rm GS} E_{\rm GS} + c^{(1)}_{\rm XS}c^{(2)}_{\rm XS} E_{\rm XS}\\
				       &=& |c^{(1)}_{\rm GS}|\sqrt{1-|c^{(1)}_{\rm GS}|^2}(E_{\rm XS}-E_{\rm GS})\label{eq_coupling_gap}
\end{eqnarray}
where the last relation can be deduced by orthonormality relations. Eq.~\ref{eq_coupling_gap} gives us the chance of expressing the coupling term as a function of the energy gap which can be calculated at different levels of approximations and a single parameter, $|c^{(1)}_{\rm  GS}|$, which is in principle unknown. We decided to calculate the gap with different multiconfigurational approaches of different accuracy and study the dependence from this parameter of $H_{12}$ and therefore of the TSCC-MoI values. Since the most sensible data are the ones close to the crossing we will focus on this regime. Furthermore, as a representative example we employed the minimal basis set since as we have seen the effect of the basis set on the gap is not particularly significant (see Table~\ref{tab_1}). In the left panel of Fig.~\ref{fig_8} we report the coupling calculated via Eq.~\ref{eq_coupling_gap} as a function of $|c^{(1)}_{\rm GS}|$. The gaps were obtained for the Be-Be distance of 2.70~{\AA} by using different RAS-SCF calculations and DMRG with the minimal basis set used in this work. As one can see, more or less independently on the method employed to calculate the gap, $H_{12}$ assumes values in the range $\pm~5~{\rm m}E_h$ with respect to the constant coupling value (see Fig.~\ref{fig_8}). Moreover, since we are analyzing the region close to the crossing, $|c^{(1)}_{\rm GS}|$ is likely to be in the range 0.4-0.8 which further narrows the possible values that $H_{12}$ can assume. We used these values to calculate the ground state energy employing the CAS-CI-TSCC-MoI. As one can see in the right panel of Fig.~\ref{fig_8}, the error introduced for the most accurate method employed (DMRG with $\chi=10^{-4}$) is smaller than $3~{\rm m}E_h$.

\subsection{Larger rings}

\begin{figure}[h]
\includegraphics[width=0.4\textwidth]{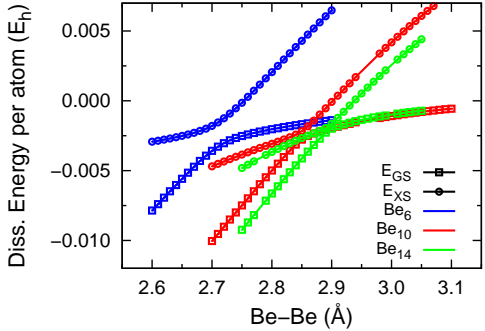}
\caption{(Color online) Dissociation curves for the ground and first excited $^1A_{1g}$ state around the avoided crossing region of Be$_6$, Be$_{10}$ and Be$_{14}$ as calculated with CCSD(T)-TSCC-MoI and the $cc$-pVDZ basis set. In order to allow an easier comparison, energies per atom are reported.}
\label{fig_9}
\end{figure}

Since the TSCC-MoI results for the Be$_{6}$ ring seemed promising, we decided to make a step further and apply the method to larger Be rings. As discussed we found a good agreement between CCSD(T)-TSCC-MoI and the MR methods applied. Therefore, in order to avoid difficulties arising from the localization of the virtual orbitals, we decided to apply CCSD(T)-TSCC-MoI to larger Be rings. Moreover, assuming that such systems present a wave function similar to Be$_6$ in the relevant electronic states, we employed the same scheme used so far based on a SA-CAS(2,2) reference. Of course, such assumption has to be taken with care since the number of emerging configurations and states grows exponentially with the system size, but in the crossing region it is reasonable that two configurations related by double excitation are of major interest as for Be$_6$. The dissociation curves of Be$_{10}$ and Be$_{14}$ obtained in such a way are compare to the data for the six-membered ring in Fig.~\ref{fig_9}. All increments up to the three-body level were introduced and $cc$-pVDZ was used once again. As one can see, smooth dissociation curves were obtained in all cases converging as expected to the same dissociation limit. The position of the crossing shifts towards larger internuclear distance going from 2.71~{\AA} for Be$_6$ to 2.89~{\AA} for Be$_{14}$. Moreover, even if the excited state results and therefore the energy gaps have to be taken with care, it is interesting that the excitation gap reduces to 4.73~m$E_h$  for Be$_{14}$ from 10.41~m$E_h$ for Be$_6$. Provided a more settled description of the excited state, the analysis of such behavior for even larger systems will be an interesting topic for future studies as it would give the chance to study energy gaps of infinite chains. 

\section{Conclusion}\label{sec_conclusion}

The two-state constant-coupling method of increments was introduced and tested on the Be$_6$ ring employing a minimal basis set and using DMRG as a benchmark. The comparison between the two methods shows encouraging results for the ground state whose dissociation curve can be described by TSCC-MoI in a size-consistent way.\\
Applying the method with the $cc$-pVDZ basis set and introducing dynamical correlation we were then able to achieve a more accurate description of the system. As discussed, the deviations from the conventional MoI are small for short and large internuclear distances, but significant around the crossing, since the standard formalism could not represent this region correctly.\\
We also discussed the validity of the constant-coupling approximation by studying the dependence of the energy on the coupling and evaluating the maximum error introduced, showing that for the system under study the approximation does not lead to an error larger than $\pm~3~{\rm m}E_h$ for the ground state.\\
The excited state obtained as a byproduct is not described in general with the same accuracy as the ground state, since it shows larger deviations from the DMRG benchmark. This was justified by discussing how the employed SA-CAS(2,2) wave function, which yields a good size-consistent description for the ground state, does not provide a proper reference for the excited state. Around the crossing, however, the employed reference seems to be appropriate and the energy difference between TSCC-MoI and DMRG is in the range $1-5$~m$E_h$ in this region. This allowed us to discuss excitation energies for Be$_6$ and larger rings.\\
Even if we treated it as a byproduct, achieving accurate description about the excited state via a local approach is an appealing perspective since it gives the chance to study excitation energies of extended and periodic systems. We will therefore address further investigations in this direction.

\acknowledgments{This research was supported by the German Research Foundation (DFG) and the Agence Nationale de la Recherche (ANR) via the project ``Quantum-chemical investigation of the metal-insulator transition in realistic low-dimensional systems'' (action ANR-11-INTB-1009 MITLOW PA1360/6-1), as well as by the Hungarian Research Fund (OTKA) under the grant number NN110360. The support of the Zentraleinrichtung f\"ur Datenverarbeitung (ZEDAT) at the Freie Universit\"at Berlin is gratefully acknowledged. EF thanks the support of the Max Planck Society via the International Max Planck Research School. At last DK and EF would like to thank PD Dr. Dirk Andrae for the many fruitful discussions and suggestions.}

\bibliography{bibliography}

\newpage

\setcounter{figure}{0}
\renewcommand\thefigure{S.\arabic{figure}}
\renewcommand\thetable{S.\Roman{table}}

\section*{Supplementary materials}

\begin{figure}[b]
\subfigure{\includegraphics[width=0.5\textwidth]{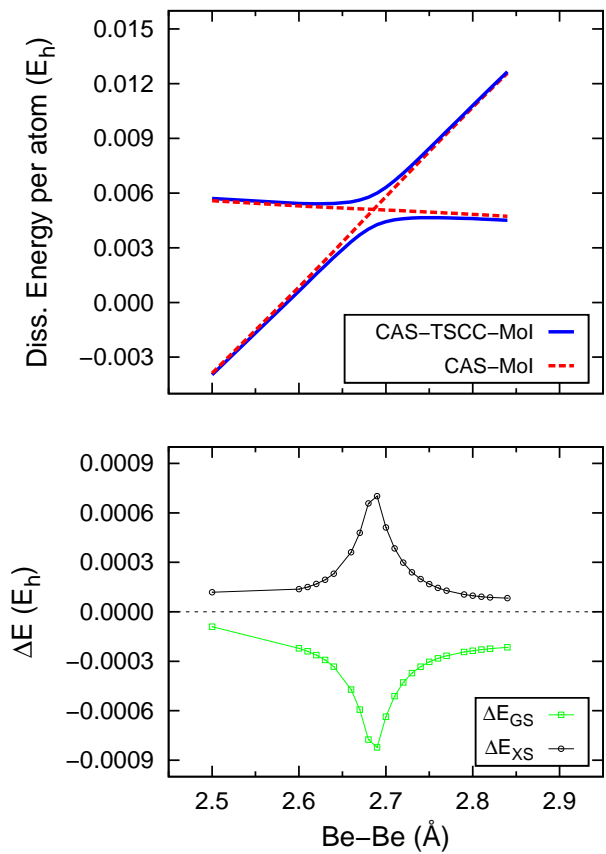}}
\caption{(Color online) Comparison between the two-state constant-coupling method of increments (TSCC-MoI) results and the ones obtained with the standard method of increments for the Be$_6$ ring. In both  cases CAS-SCF was employed at the three-body level with the $cc$-pVDZ basis set. The reported values are energies per atom.}
\label{fig_sm_1}
\end{figure}

\begin{figure}
\subfigure{\includegraphics[width=0.5\textwidth]{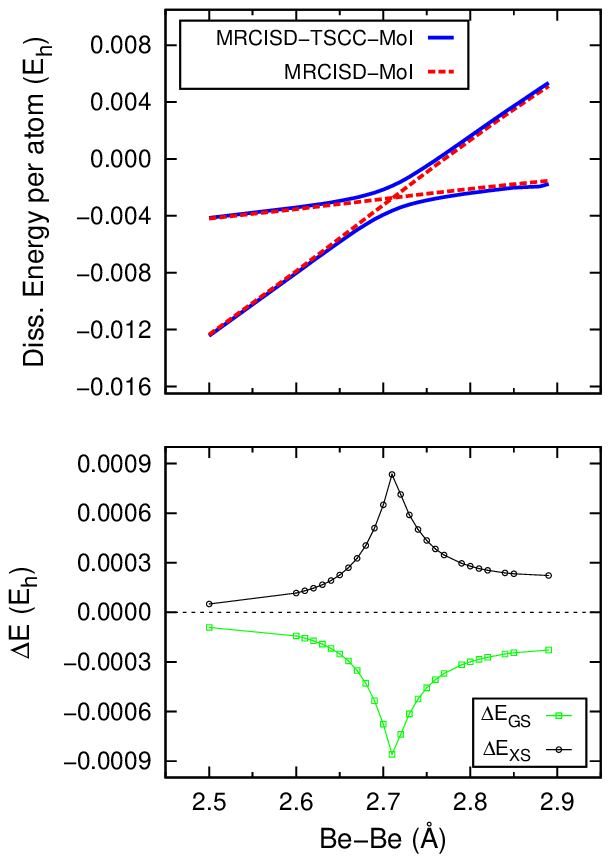}}
\caption{(Color online) Comparison between the two-state constant-coupling method of increments (TSCC-MoI) results and the ones obtained with the standard method of increments for the Be$_6$ ring. In both  cases MRCISD was employed at the three-body level with the $cc$-pVDZ basis set. The reported values are energies per atom.}
\label{fig_sm_2}
\end{figure}

\begin{figure}
\subfigure{\includegraphics[width=0.5\textwidth]{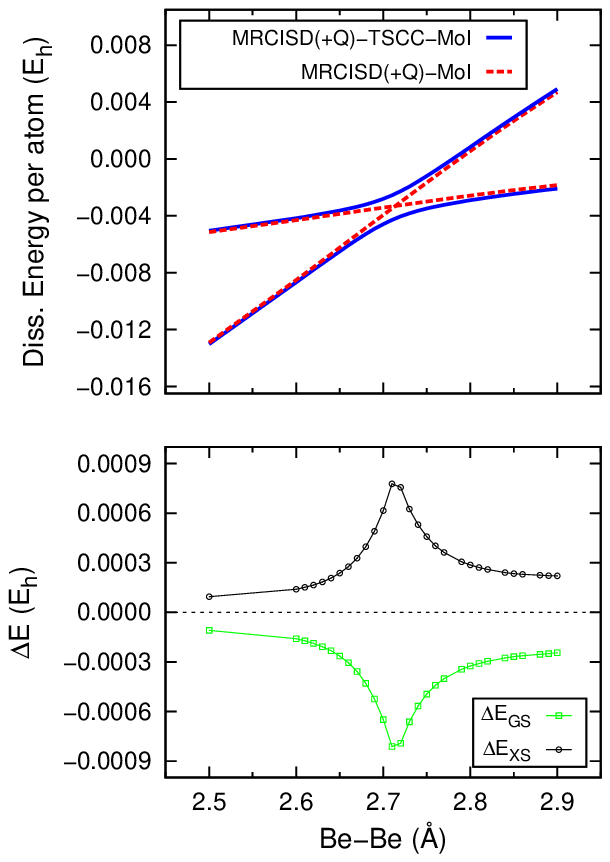}}
\caption{(Color online) Comparison between the two-state constant-coupling method of increments (TSCC-MoI) results and the ones obtained with the standard method of increments for the Be$_6$ ring. In both  cases MRCISD(+Q) was employed at the three-body level with the $cc$-pVDZ basis set. The reported values are energies per atom.}
\label{fig_sm_3}
\end{figure}

In Fig.~\ref{fig_sm_1},~\ref{fig_sm_2}~and~\ref{fig_sm_3} the results obtained with the conventional MoI and the TSCC-MoI along with CAS-SCF, MRCISD and MRCISD(+Q) are shown.
For a better comparison the differences between the ground state of the TSCC-MoI and the lower MoI energies as well as between the excited state of the TSCC-MoI and the higher MoI energies are shown below the dissociation curves for every method.

\end{document}